# HOLOGRAPHY, QUANTUM GEOMETRY, AND QUANTUM INFORMATION THEORY

P. A. Zizzi

Dipartimento di Astronomia dell'Università di Padova, Vicolo dell'Osservatorio, 5

35122 Padova, Italy

e-mail: zizzi@pd.astro.it

ABSTRACT

We interpret the Holographic Conjecture in terms of quantum bits (qubits). N-qubit states are associated with surfaces that are punctured in N points by spin networks' edges labelled by the spin-$\frac{1}{2}$ representation of $SU(2)$, which are in a superposed quantum state of spin "up" and spin "down". The formalism is applied in particular to de Sitter horizons, and leads to a picture of the early inflationary universe in terms of quantum computation. A discrete micro-causality emerges, where the time parameter is being defined by the discrete increase of entropy.

Then, the model is analysed in the framework of the theory of presheaves (varying sets on a causal set) and we get a quantum history. A (bosonic) Fock space of the whole history is considered. The Fock space wavefunction, which resembles a Bose-Einstein condensate, undergoes decoherence at the end of inflation. This fact seems to be responsible for the rather low entropy of our universe.

# 1  INTRODUCTION

Today, the main challenge of theoretical physics is to settle a theory of quantum gravity that will reconcile General Relativity with Quantum Mechanics.

There are three main approaches to quantum gravity: the canonical approach, the histories approach, and string theory. In what follows, we will focus mainly on the canonical approach; however, we will consider also quantum histories in the context of topos theory [1-2].

A novel interest in the canonical approach emerged, about ten years ago, when Ashtekar introduced his formalism [3] which lead to the theory of loop quantum gravity [4].

In loop quantum gravity, non-perturbative techniques have led to a picture of quantum geometry. In the non-perturbative approach, there is no background metric, but only a bare manifold. It follows that at the Planck scale geometry is rather of a polymer type, and geometrical observables like area and volume have discrete spectra.

Spin networks are relevant for quantum geometry. They were invented by Penrose [5] in order to approach a drastic change in the concept of space-time, going from that of a smooth manifold to that of a discrete, purely combinatorial structure. Then, spin networks were re-discovered by Rovelli and Smolin [6] in the context of loop quantum gravity. Basically, spin networks are graphs embedded in 3-space, with edges labelled by spins $j=0, 1/2, 1, 3/2$...and vertices labelled by intertwining operators. In loop quantum gravity, spin networks are eigenstates of the area and volume-operators [7].

However, this theory of quantum geometry, does not reproduce classical General Relativity in the continuum limit. For this reason, Reisenberg and Rovelli [8] proposed that the dynamics of spin networks could be described in terms of spacetime histories, to overcome the difficulties of the canonical approach. In this context, the spacetime histories are represented by discrete combinatorial structures that can be visualised as triangulations. The merging of dynamical triangulations with topological quantum field theory (TQFT) has given rise to models of quantum gravity called "spin foam models" [9] which seem to have continuum limits.

Anyway, as spin foam models are Euclidean, they are not suitable to recover causality at the Planck scale. For this purpose, the theory should be intrinsically Lorentzian. However, the very concept of causality becomes uncertain at the Planck scale, when the metric undergoes quantum fluctuations as Penrose [10] argued. So, one should consider a discrete alternative to the Lorentzian metric, which is the causal set (a partially ordered set-or poset- whose elements are events of a discrete space-time). Such theories of quantum gravity based on the casual set were formulated by Sorkin et al. [11].

Rather recently, a further effort in trying to recover causality at the Planck scale, has been undertaken by Markopoulou and Smolin [12]. They considered the evolution of spin networks in discrete time steps, and they claimed that the evolution is causal because the history of evolving spin networks is a causal set. However, it is not clear yet whether causality can really be achieved at the Planck scale, at least in the context of causal sets.

In this paper, we also consider the issue of causality at the Planck scale in the framework of causal sets, although our results do not promote this aspect as decisively as in [12]. In our opinion, the very concept of causality makes sense in the quasi-classical limit, in relation with the fact that an observer is needed, and this observer cannot "stand" on the event at the Planck scale.

Moreover, our picture is related to some other issues, mainly the holographic conjecture [13-14] and quantum information theory.

The issue of quantum information started with Neumann [15], who gave a mathematical expression of quantum entropy in photons and other particles subjected to the laws of quantum mechanics. However, Neumann's entropy lacked a clear interpretation in terms of information theory. It was Schumacher [16], who showed that Neumann's entropy has indeed a related meaning. Moreover, Holevo [17], and Levitin [18], found that the value of quantum entropy is the upper limit of the amount of quantum information that can be recovered from a quantum particle (or from a group of particles).

Quantum information differs from classical information [19], in several aspects.

The elementary unit of classical information is the bit, which is in one of the two states "true"=1 and "false"=0, and obeys Boolean logic. A classical bit is contextual-free i.e. it does not depend on what other information is present. Finally, a classical bit can be perfectly copied.

In quantum information, the elementary unit is the quantum bit (or qubit), a coherent superposition of the two basis states 0 and 1. The underlying logic is non-Boolean. Different from classical bits, a qubit is contextual [20], and cannot be copied, or "cloned" [21].

The emerging fields of quantum computation [22], quantum communication and quantum cryptography [23], quantum dense coding [24], and quantum teleportation [25], are all based on quantum information theory. Moreover, quantum information theory is expected to illuminate some conceptual aspects of the foundations of Quantum Mechanics.

In this paper, we shall illustrate, in particular, the interconnections between quantum information theory and quantum gravity.

As spin networks are purely mathematical entities, they do not carry any information of their own because information is physical [26]. However, they do carry information when they puncture a surface transversely. Each puncture contributes to a pixel of area (a pixel is one unit of Planck area) and, because of the holographic principle, this pixel will encode one unit of classical information (a bit).

The starting idea of our paper is that a surface can be punctured by a spin network's edge (labelled by the j=1/2 representation of SU (2)), which is in a superposed quantum state of spin "up" and spin "down". This induces the pixel to be in the superposed quantum state of "on" and "off", which encodes one unit of quantum information, or qubit. By the use of the Bekenstein bound [27], we show that the entropy of an N-punctured surface is in fact the entropy of an N-qubit.

Moreover, because of the Holevo-Levitin theorem [17-18], our picture provides a time parameter in terms of a discrete increase of entropy. Thus the time parameter is discrete, and quantized in Planck's units. This is in agreement with Penrose's argument [28], that a theory of quantum gravity should be time-asymmetric. However, our result is not strong enough to claim that causality is actually being recovered at the Planck scale. In fact, the thermodynamic arrow of time is related to the psychological arrow of time, and at the Planck scale, the latter is missing because there are no observers. This is not just a matter of philosophy, but it is a real handicap.

The "events" of our causal set are not just points of a discrete spacetime, but they are the elements of a poset whose basic set is the set of qubits and whose order relation follows directly from the quantum entropy of the qubits.

Events that are not "causally" related (qubits with the same entropy) form space-like surfaces (or antichains).

The use of topos theory (in particular the theory of presheaves [29]) makes it possible to attach an Hilbert space to each event, and to build up discrete evolution operators [30], between different Hilbert spaces. In this way, we get a quantum history, which turns out to be a quantum information interpretation of the theory of inflation.

However, as we already pointed out, we find an internal contradiction, in considering causality in the framework of causal sets: an observer seems to be required for consistency.

In fact presheaves (or varying sets on a causal set) obey the Heyting algebra which implies the intuitionistic logic [31-32], which in turn is related to the concept of time flow. One can then imagine a Boolean-minded observer who has to move in time [33], in order to grasp the underlying quantum logic of the universe. Obviously, this picture has no meaning at the Planck time, when there was no observer at all. The above picture strictly depends on the fact that in the theory of presheaves we consider Boolean sub-lattices of the quantum lattice. To drop the Boolean observer, one should discard Boolean sub-lattices and consider the entire quantum lattice as a whole, endowed with a non-Boolean logic. However, we cannot get the Hilbert space of the entire history, because of the existence of unitary evolution operators among different Hilbert spaces.

Thus, in order to escape the "problem of the observer" we perform the tensor sum of all the Hilbert spaces attached to the events of the causal set, and we get a (bosonic) Fock space.

Although we find that the logic associated with the Fock space is Boolean, we are able to eliminate the observer at the Planck scale by depriving him of time evolution. In fact, the Fock space wave function $|\Psi\rangle$, which is the coherent quantum superposition of all the events (qubits), leads to an a-temporal picture of the early universe.

The wavefunction $|\Psi\rangle$ is the coherent quantum state of multiple bosonic qubits, and resembles a Bose-Einstein condensate [34]. It can maintain coherence as far as thermal noise is absent, that is, as far as inflation is running: a cosmological era when the universe is cool and vacuum-dominated. We find that the coherent quantum state decoheres at the end of inflation, giving rise to the (rather) low entropy of our universe. In fact, the present entropy seems to be related to the quantum information stored in $|\Psi\rangle$, which was lost to the emerging environment when thermal decoherence took place [35].

Only after $|\Psi\rangle$ has decohered, one can in principle retrodict the quantum past [36], "as" it was recorded by some ancient observer. It seems to us that only in this sense causality can be "restored" at the Planck scale.

This paper is organised as follows.

In Sect.2, we review the holographic conjecture by 't Hooft and Susskind, and we interpret the information stored on a boundary surface in terms of quantum bits rather than classical bits.

In Sect.3, we illustrate the relation between spin networks that puncture a boundary surface, and the qubits stored on the surface.

In Sect.4, we put forward the mathematical formalism. We select the subspace of unentangled symmetric qubits of the full $2^N$-dimensional Hilbert space of N qubits. We find that the symmetric 1-qubit acts as a creation operator in this subspace. Then the formalism is applied to de Sitter horizons, which are interpreted as qubits. This leads to an interpretation of inflation in terms of quantum information.

In Sect.5, we show how the vacuum (the classical bit) at the unphysical time t=0, evolved to the 1-qubit at the Planck time by passing through a quantum logic gate.

In Sect.6, we show that in this model a discrete micro-causality emerges, where the time parameter is given in terms of the discrete increase of entropy.

In Sect.7, we formalise our model within the framework of the theory of presheaves. Qubits are interpreted as events of a causal set, where the order relation is given in terms of the entropy. We get thus a quantum history which is the ensemble of all the finite-dimensional Hilbert spaces of N-qubits. We find that the wavefunction of the entire history resembles a Bose-Einstein condensate.

In Sect.8, we consider decoherence of the Bose-Einstein condensate. We find that the rather low value of the entropy of our present universe can be achieved only if the wavefunction collapsed at the end of inflation.

In Sect.9, we interpret the N-qubits as N quantum harmonic oscillators. This allows us to find the discrete energy spectrum, and in particular, the energy at the time when inflation ended. This energy is interpreted as the reheating temperature.

Sect. 10 is devoted to some concluding remarks.

## 2   HOLOGRAPHY AND QUANTUM BITS

The Holographic Conjecture of 't Hooft [13], and Susskind [14], is based on the Bekenstein bound [27]:

(2.1)   $S = \dfrac{A}{4}$

where S is the entropy of a region of space of volume V, and A is the area, in Planck units, of the surface bounding V.

The entropy S of a quantum system is equal to the logaritm of the total number $N$ of degrees of freedom, i.e. the dimension of the Hilbert space:

(2.2)   $S = \ln N = \ln(\dim H)$

In a discrete theory of N spins that can take only two values (Boolean variables) the dimension of the Hilbert space is $2^N$, hence the entropy directly counts the number of Boolean degrees of freedom:

(2.3) $\quad S = N \ln 2$

From eqs. (2.1) and (2.3) one gets:

(2.4) $\quad N = \dfrac{A}{4 \ln 2}$

In a region of space-time surrounding a black hole, the number of Boolean degrees of freedom is proportional to the horizon area A.

't Hooft proposed that it must be possible to describe all phenomena within the bulk of a region of space of volume V by a set of degrees of freedom which reside on the boundary, and that this number should not be larger than one binary degree of freedom per Planck area.

All this can be interpreted as follows: each unit of Planck area (a pixel) is associated with a classical bit of information. The bit is the elementary quantity of information, which can take on one of two values, usually 0 and 1.

Therefore, any physical realization of a bit needs a system with two well defined states.

Thus, in the classical picture of holography, a pixel is a two-state system which realizes a classical bit.

In fact a pixel can be either $"on" \equiv 1$ or $"off" \equiv 0$, where we take the convention that the pixel is $"on"$ when it is punctured by the edge of a spin network [6], (in the spin-$\frac{1}{2}$ representation of SU(2)) in the state $\left|+\frac{1}{2}\right\rangle$ and that the pixel is $"off"$ when it is punctured by a spin network's edge in the state $\left|-\frac{1}{2}\right\rangle$.

But a pixel can be $"on"$ and $"off"$ at the same time, if it is punctured by a (open) spin nework's edge in the superposed quantum states:

(2.5) $\quad \frac{1}{\sqrt{2}}\left( \left| \frac{1}{2} \right\rangle \pm \left| -\frac{1}{2} \right\rangle \right).$

Let us represent the states $\left| +\frac{1}{2} \right\rangle$ and $\left| -\frac{1}{2} \right\rangle$ as the vectors $\begin{pmatrix} 1 \\ 0 \end{pmatrix}$ and $\begin{pmatrix} 0 \\ 1 \end{pmatrix}$ respectively.

The superposed quantum states in eq. (2.5) can be obtained by the action of the unitary operator:

(2.6) $\quad U_{\sigma_2} = \frac{1}{\sqrt{2}}(1 + i\sigma_2)$

on the states $\left| -\frac{1}{2} \right\rangle$ and $\left| +\frac{1}{2} \right\rangle$ respectively.

where $\sigma_2$ is the Pauli matrix: $\sigma_2 = \begin{pmatrix} 0 & -i \\ i & 0 \end{pmatrix}$

In this case, the pixel is associated with a quantum bit of information (or qubit).

A quantum bit differs from the classical bit in so far as it can be in both states $|0\rangle$ and $|1\rangle$ at the same time: the single qubit being written then as a superposition:

(2.7) $\quad a|0\rangle + b|1\rangle$

where a and b are the complex amplitudes of the two states, with the condition:

(2.8) $\quad |a|^2 + |b|^2 = 1$

The relation between the qubit in eq. (2.7) and the superposed quantum state in eq. (2.5) lies in the fact that the group manifold of $SU(2)$ can be parametrized by a 3-sphere with unit radius.

In fact, the most general form of a $2 \times 2$ unitary matrices of unit determinant is:

$$U = \begin{pmatrix} a & b \\ -b^* & a^* \end{pmatrix} \qquad |a|^2 + |b|^2 = 1$$

where a and b are complex numbers.

Of course, this interpretation of a pixel as a qubit state is in contraddiction with classical Boolean logic where intermediate states between $|0\rangle$ and $|1\rangle$ are not possible.

The question is: What changes in the holographic picture, when one choses to associate a pixel with a quantum bit of information instead of a classical bit?

The pixel is not anymore a classical unit of surface area: it is itself a quantum state.

However, the dimension of the Hilbert space does not change: it is still $2^N$ for a quantum system of N qubits.

What changes is the interpretation: one goes from a classical Boolean logic to a quantum (non-Boolean) logic.

While Boolean logic is distributive, non-Boolean logic (Quantum logic) is nondistributive:

$a \wedge (b \vee c) \neq (a \wedge b) \vee (a \wedge c)$

In what follows, qubit states will be associated with boundary surfaces whose areas are given in units of the Planck area.

Surfaces with area equal to N pixels will be interpreted as N-qubit states.

## 3  QUANTUM INFORMATION VERSUS QUANTUM GRAVITY

When one considers N qubits, the dimensionality of the Hilbert space grows as $2^N$.

We will call these Hilbert spaces $H_N$. The basis states are the $2^N$ strings of lenght N (for example:

$|0,0\rangle$  $|0,1\rangle$  $|1,0\rangle$  $|1,1\rangle$ for N=2).

There is a link between qubits and spin networks in the context of the Holographic principle.

In loop quantum gravity, non-perturbative techniques have led to a quantum theory of geometry in which operators corresponding to lengths, area and volume have discrete spectra.

Of particular interest are the spin network states associated with graphs embedded in 3-space with edges labelled by spins

$$j = 0, \frac{1}{2}, 1, \frac{3}{2}, \ldots$$

and vertices labeled by intertwining operators.

If a single edge punctures a 2-surface transversely, it contributes an area proportional to:

(3.1) $\quad l^{*2} \sqrt{j(j+1)}$

where $l^*$ is the Planck length.

The points where the edges end on the surface are called "punctures".

If the surface is punctured in n points, the area is proportional to:

(3.2) $\quad l^{*2} \sum_n j_n(j_n + 1)$

Hence, gravity at the Planck scale is organized into branching flux tubes of *area*.

What happens when this picture is applied to a black hole horizon? The flux lines pierce the black hole horizon and excite curvature degrees of freedom on the surface [37]. These excitations are described by Chern-Simons theory and account for the black hole entropy.

The very important feature of Chern-Simons theory is that is possible to "count" the number of states: for a large number of punctures, the dimension of the Hilbert space $H_P$ for a permissible set P of punctures $P = \{j_{p_1}, \ldots j_{p_n}\}$ goes as:

(3.3) $\quad \dim H_P \approx \prod_{j_p \in P} (2j_p + 1)$

The entropy of the black hole will be then:

(3.4) $\quad S = \ln \sum_P \dim H_P = const \frac{A}{4l^{*2}\gamma}$

where A is the area of the horizon, $l^*$ is the Planck length and $\gamma$ is a parameter of the theory called the Immirzi parameter [38].

Then, the best realization of the holographic hypothesis seems to be a topological quantum field theory like Chern-Simons. It should be noted that the states which dominate the counting of degrees of freedom correspond to punctures labelled by $j = \frac{1}{2}$. This fact reminds us of Wheeler's picture

"it from Bit" [39], of the origin of black hole entropy.

In some sense, we could formulate instead "it from Qubit".

In what follows, N-punctured surfaces will be associated with N qubits. Thus, the merging of quantum information theory with quantum geometry and Chern-Simons theory will illuminate the quantum features of holography.

In particular, we will consider de Sitter-like horizons of area [40]:

(3.5) $\quad A_n = (n+1)^2 L^{*2}$

at times:

(3.6) $\quad t_n = (n+1) t^*$ with: n=0,1,2,3...

where $L^*$ and $t^*$ are the Planck lenght and time, respectively.

In what follows we will denote with n=0,1,2... the label of area eigenvalues of the horizons, and with N=1,2,3.....the number of pixels and associated qubits.

Let us consider now spin networks with edges labelled by the spin-1/2 representation of SU(2), which puncture the de Sitter horizons in the points $p_1, p_2, ....p_N$ with:

(3.7) $\quad N = (n+1)^2$

For n=0, (N=1) there will be only one puncture, $p_1$, giving rise to the minimal area, $A_0 = L^{*2}$ (one pixel). In this case, the de Sitter horizon coincides with the absolute horizon of an Euclidean Planckian black hole [40].

For N=1 we get:

(3.8) $\quad \dim H_1 = 2$

(3.9) $\quad S = \ln 2 = \frac{const}{4\gamma}$

where we have taken $A = l^{*2}$, as a Planckian black hole carries one unit of quantum information (its horizon is a 1-qubit state).

The basis states of $H_1$ are $|on\rangle$ and $|off\rangle$:

(3.10) $\langle on \| on \rangle = \langle off \| off \rangle = 1$

(3.11) $\langle on \| off \rangle = 0$

The one-pixel state can be written as a superposition of the basis states. For future convenience, we will consider the maximally coherent state, i.e, an equal superposition of $|on\rangle$ and $|off\rangle$:

(3.12) $|1\rangle = \frac{1}{\sqrt{2}}(|on\rangle \pm |off\rangle)$

# 4 THE MATHEMATICAL FORMALISM

Let us call $|Q_1\rangle$ the most general 1-qubit state:

(4.1) $|Q_1\rangle = a|on\rangle + b|off\rangle$

with: $|a|^2 + |b|^2 = 1$

The maximally coherent 1-qubit states (equal superpositions of $|on\rangle$ and $|off\rangle$) are:

(4.2) $|1\rangle^S = \frac{1}{\sqrt{2}}(|on\rangle + |off\rangle)$

(4.3) $|1\rangle^A = \frac{1}{\sqrt{2}}(|on\rangle - |off\rangle)$

where A and S stand for "symmetric" and "antisymmetric" respectively.

The most general 2-qubit states are:

(4.4) $|Q_2\rangle = \sum_{m,n} \sum_{j}^{K(m,n)} c_{mn}^j |mn\rangle^j$

where:

m=number of "on", n=number of "off" and n+m=2

K(0,2)=K(2,0)=1

K(1,1)=2

The unentangled basis for 2-qubits is:

(4.5) $\quad |on,on\rangle, |off,off\rangle, |on,off\rangle, |off,on\rangle$

The entangled basis for 2-qubits (Bell states, maximally entangled) is:

(4.6) $\quad |\Phi_\pm\rangle = \frac{1}{\sqrt{2}}(|on,on\rangle \pm |off,off\rangle)$

(4.7) $\quad |\Psi_\pm\rangle = \frac{1}{\sqrt{2}}(|on,off\rangle \pm |off,on\rangle)$

Unentangled 2-qubits (product states of two 1-qubits) are:

(4.8) $\quad |2\rangle_1^S = \frac{1}{2}(|on,on\rangle + |on,off\rangle + |off,on\rangle + |off,off\rangle) = |1\rangle^S |1\rangle^S$

(4.9) $\quad |2\rangle_2^S = \frac{1}{2}(|on,on\rangle - |on,off\rangle - |off,on\rangle + |off,off\rangle) = |1\rangle^A |1\rangle^A$

(4.10) $\quad |2\rangle_1^A = \frac{1}{2}(|on,on\rangle - |on,off\rangle + |off,on\rangle - |off,off\rangle) = |1\rangle^S |1\rangle^A$

(4.11) $\quad |2\rangle_2^A = \frac{1}{2}(|on,on\rangle + |on,off\rangle - |off,on\rangle - |off,off\rangle) = |1\rangle^A |1\rangle^S$

These 4 states can be written in the Bell basis as:

(4.12) $\quad |2\rangle_{1,2}^S = \frac{1}{\sqrt{2}}(|\Phi_+\rangle \pm |\Psi_+\rangle)$

(4.13) $\quad |2\rangle_{1,2}^A = \frac{1}{\sqrt{2}}(|\Phi_-\rangle \mp |\Psi_-\rangle)$

The unentangled 2-qubits $|2\rangle_{1,2}^S$ and $|2\rangle_{1,2}^A$ are given by eq. (4.4) in the case $c_j = c_k$. In particular, for the states $|2\rangle^S$ the $c_{mn}$ are symmetric under permutation of m and n, while, for the states $|2\rangle^A$, they are antisymmetric.

So, general unentangled 2-qubits can be written as:

$$(4.14) \quad |Q_2\rangle^{UN} = \sum_{mn} \sum_{j=1}^{4} c_{mn}^{j} |mn\rangle^{j} \quad \text{with m+n=2, and all } c_j = c_k$$

where "UN" stands for "unentangled".

The entangled 2-qubits are of two kinds:

$$(4.15) \quad |Q_2\rangle^{E}_{(1)} = \sum_{mn} \sum_{j=1}^{4} c_{mn}^{j} |mn\rangle^{j} \quad \text{with m+n=2, and not all } c_j = c_k$$

$$(4.16) \quad |Q_2\rangle^{E}_{(2)} = \sum_{mn} \sum_{j=1}^{M<4} c_{mn}^{j} |mn\rangle^{j} \quad \text{with m+n=2}$$

where "E" stands for "entangled".

We shall consider the action of $|1\rangle^{S}$ (creation operator) on the 2-qubits.

I) On the entangled 2-qubits:

$$(4.17) \quad |1\rangle^{S} \otimes |Q_2\rangle^{E}_{(1)} = \sum_{m'n'} \sum_{j=1}^{8} c_{m'n'}^{j} |m'n'\rangle^{j} \equiv |Q_3\rangle^{E}_{(1)} \quad \text{with m'+n'=N+1=3 and not all } c_j = c_k$$

$$(4.18) \quad |1\rangle^{S} \otimes |Q_2\rangle^{E}_{(2)} = \sum_{m'n'} \sum_{j=1}^{M'<8} c_{m'n'}^{j} |m'n'\rangle^{j} \equiv |Q_3\rangle^{E}_{(2)}$$

II) - On the unentangled 2-qubits:

$$(4.19) \quad |1\rangle^{S} \otimes |Q_2\rangle^{UN} = \sum_{m'n'} \sum_{j=1}^{8} c_{m'n'}^{j} |m'n'\rangle^{j} \equiv |Q_3\rangle^{UN} \quad \text{with m'+n'=3 and all } c_j = c_k$$

In what follows, we will consider only the unentangled 2-qubits $|2\rangle^{S}_{1}$ (and we will call them simply $|2\rangle$) for which it holds:

$$(4.20) \quad |1\rangle^{S} \otimes |2\rangle = |3\rangle \quad \text{where:}$$

$$|3\rangle = \frac{1}{2\sqrt{2}} (|on,on,on\rangle + |on,on,off\rangle + |on,off,on\rangle + |on,off,off\rangle + |off,on,on\rangle + |off,on,off\rangle +$$
$$+ |off,on,off\rangle + |off,off,on\rangle + |off,off,off\rangle)$$

The entangled N-qubits are:

$$(4.21) \quad |Q_N\rangle^{(E)}{}_{(1)} = \sum_{mn} \sum_{j=1}^{M=2^N} c_{mn}^j |mn\rangle^j \quad \text{with m+n=N and \underline{not all} } c_j = c_k$$

$$(4.22) \quad |Q_N\rangle^{(E)}{}_{(2)} = \sum_{mn} \sum_{j=1}^{M<2^N} c_{mn}^j |mn\rangle^j \quad \text{with m+n=N}$$

and the unentangled N-qubits:

$$(4.23) \quad |Q_N\rangle^{(UN)} = \sum_{mn} \sum_{j=1}^{M=2^N} c_{mn}^j |mn\rangle^j \quad \text{with m+n=N and \underline{all} } c_j = c_k$$

A subset of the $|Q_N\rangle^{(UN)}$ is the set of the symmetric states:

$$(4.24) \quad |Q_N\rangle^{(UN.Symm)} = \sum_{mn} \sum_{j=1}^{M=2^N} c_{mn}^j |mn\rangle^j \quad \text{where m+n=N, \underline{all,} } c_j = c_k \text{ and the } c_{mn} \text{ are symmetric}$$

under permutation of m and n.

Finally, a particular subset of the $|Q_N\rangle^{(UN.Symm)}$ is the set of states:

$$(4.25) \quad |N\rangle = \frac{1}{\sqrt{2}^N} |1\rangle^{(Symm)\otimes N}$$

The action of $|1\rangle^{(S)}$ on the N-qubits is:

$$(4.26) \quad |1\rangle^{(S)} \otimes |Q_N\rangle^{(E)}_{(1)} = |Q_{N+1}\rangle^{(E)}_{(1)}$$

$$(4.27) \quad |1\rangle^{(S)} \otimes |Q_N\rangle^{(E)}_{(2)} = |Q_{N+1}\rangle^{(E)}_{(2)}$$

$$(4.28) \quad |1\rangle^{(S)} \otimes |Q_N\rangle^{(UN)} = |Q_{N+1}\rangle^{(UN)}$$

So, in general, the action of $|1\rangle^{(S)}$ on a N-qubit whatever entangled or unentangled is:

$$(4.29) \quad |Q_N\rangle = \sum_{mn} \sum_{j=1}^{M\leq 2^N} c_{mn}^j |mn\rangle^j \quad \text{with m+n=N, is:}$$

$$(4.30) \quad |1\rangle^{(S)} \otimes |Q_N\rangle = |Q_{N+1}\rangle$$

where:

(4.31) $|Q_{N+1}\rangle = \sum_{m'n'} \sum_{j=1}^{M' \leq 2^{N'}} c_{m'n'}^{j} |m'n'\rangle^{j}$ with m'+n'=N'=N+1

In particular, the action of $|1\rangle^{(S)}$ on the qubits $|N\rangle$ is the action of a creation operator on a N-particles state is:

(4.32) $|1\rangle^{(S)} \otimes |N\rangle = |N+1\rangle$

We will use the following notations:

a)-The kets $|N\rangle$:

(4.33) $|0\rangle = 1$

Note that N=0 corresponds to $n = -1$, that is, to $t_{-1} = 0$. We will not attach any physical meaning to this time $t_{-1}$ as we are discarding all times preceding the Planck time $t_0 = t^*$.

The state $|0\rangle$ that will be called the vacuum state, is the basis state of an Hilbert space $H_0$ of dimensionality 1, but it is not a proper qubit state.

So, the quantum pixel states will be labelled by N=1,2,3....

(4.34) $|1\rangle = \frac{1}{\sqrt{2}}(|on\rangle + |off\rangle)$

(This state, which is the one-pixel state associated with one qubit of information, describes the quantum state of the first de Sitter horizon (n=0) at the Planck time $t_0 = t^*$; in fact, it is the horizon of a Planckian black hole).

$|2\rangle = \frac{1}{2}(|on,on\rangle + |on,off\rangle + |off,on\rangle + |off,off\rangle)$

$|3\rangle = \frac{1}{2\sqrt{2}}(|on,on,on\rangle + |on,on,off\rangle + |on,off,on\rangle + |on,off,off\rangle + |off,on,on\rangle +$

$+ |off,on,off\rangle + |off,off,on\rangle + |off,off,off\rangle)$

and so on..

In general, the N-pixel state is given by:

(4.35) $|N\rangle = \frac{1}{\sqrt{2}^N}|1\rangle^N$

They are vectors in a symmetric subspace $H_N^{(S)}$ of he $2^N$-dimensional Hilbert space $H_N$.

$H_N^{(S)}$ is the tensor product: $H_N^{(S)} = H_1^{(s)\otimes N}$.

Notice that, in the full Hilbert space, there would be also entangled states.

The qubits $|N\rangle$, which are product of states, are disentangled by definition. Entanglement is due to the application of operations which act on two qubits simultaneously, i.e, they have to interact in a nonlinear way, quite unlike the Holographic picture, where everything is linear.

It appears as if the state $|1\rangle$ would act as a creation operator. Indeed, the state $|1\rangle$ makes jumps from a $2^N$-dimensional Hilbert space $H_N$ to a $2^{N+1}$-dimensional Hilbert space $H_{N+1}$:

(4.36) $|1\rangle|N\rangle = |N+1\rangle$

where $|N\rangle \in H_N$ and $|N+1\rangle \in H_{N+1}$

**b)-The bra $\langle N|$**

whose products with the kets $|M\rangle$ are defined by the following relations:

$\langle A|(|B_1\rangle \times |B_2\rangle \times ...... |B_N\rangle) = \frac{1}{N}(\langle A\|B_1\rangle|B_2\rangle \times ....... |B_N\rangle + ......\langle A\|B_N\rangle|B_1\rangle \times ...... |B_{N-1}\rangle)$

With these conventions:

(4.37) $\langle N|M\rangle = |M - N\rangle$    for N<M

$\quad\quad\quad = 0$    for N>M

The bra $\langle 1|$ acts as a annihilation operator:

(4.38) $\langle 1\|N\rangle = |N-1\rangle$

(4.39) $\langle 1\|0\rangle = 0$

**c)-The creation and annihilation operators:**

(4.40) $\quad a^+ = |1\rangle; \quad a = \langle 1|$

such that:

(4.41) $\quad a^+|N\rangle = |N+1\rangle$

(4.42) $\quad a|N\rangle = |N-1\rangle$

(4.43) $\quad a|0\rangle = 0$

Having normalized the scalar product to unity: $\langle 1|1\rangle = 1$

we get:

(4.44) $\quad aa^+ = 1$

Let us define the hermitian operator:

(4.45) $\quad P_1 = a^+a \equiv |1\rangle\langle 1|$

as, by the use of eq.(4.36) it holds:

(4.46) $\quad a^+aa^+a = a^+a$

$P_1$ satisfies the relation:

(4.47) $\quad P_1^2 = P_1$

Thus, $P_1$ is a projection operator by definition.

Also, $P_1$ is a number operator with only two eigenvalues (0 and 1):

(4.48) $\quad P_1|0\rangle = 0; \quad P_1|N\rangle = |N\rangle \quad$ (for N=1,2,3...)

**d)-The de Sitter horizon states $|\Psi(t_n)\rangle$:**

A N-qubit state $|N\rangle$ can be attached to each de Sitter horizon of area $A_n = (n+1)^2 L^{*2}$ at time $t_n = (n+1)t^*$:

(4.49) $\quad |\Psi(t_0)\rangle = |1\rangle$

(4.50) $|\Psi(t_1)\rangle = |4\rangle = |1\rangle^4$

(4.51) $|\Psi(t_2)\rangle = |9\rangle = |1\rangle^9$

(4.52) $|\Psi(t_3)\rangle = |16\rangle = |1\rangle^{16}$

and so on. In general:

(4.53) $|\Psi(t_n)\rangle = |(n+1)^2\rangle = |1\rangle^{(n+1)^2}$

where the state of the system at time $t = t_n$ is given by:

(4.54) $|\Psi(t_n)\rangle = |1\rangle^{n(n+2)} |\Psi(t_0)\rangle$

where $|\Psi(t_0)\rangle \equiv |1\rangle$ is the initial state at the Planck time $t_0 = t^*$.

## 5 THE VACUUM STATE AND THE QUANTUM LOGIC GATE

The vacuum state $|0\rangle = 1$ is the basis state of an Hilbert space $H_0$ of dimension 1. As we already mentioned, this state is not a quantum bit, but just a classical bit, i.e, the elementary classical quantity of information which can take on one of two values, usually "0" and "1" (in our case "on" and "off").

This means that the vacuum state is either "on" or "off", so the vacuum obeys a Boolean logic, not a Quantum logic (in which case it could be "on" or "off", i.e, it would be a qubit).

The vacuum is then a pixel which is either "on" (punctured by a spin network's edge in the state $\left|+\frac{1}{2}\right\rangle$) or "off" (punctured by a spin network's edge in the state $\left|-\frac{1}{2}\right\rangle$).

How can the vacuum state (the classical bit) at the "unphysical time": $t_{-1} = 0$ "evolve" to the state

$|1\rangle = \frac{1}{\sqrt{2}}(|on\rangle + |off\rangle)$ (the quantum bit) at the Planck time $t_0 = t^*$?

A possible answer is that the vacuum state "passes" through a *quantum logic gate*.

Logic gates may be constructed for quantum bits. We shall consider only one-bit unitary operations.

Let us consider a single quantum bit.

If we represent the basis $|on\rangle$ and $|off\rangle$ as the vectors $\begin{pmatrix}1\\0\end{pmatrix}$ and $\begin{pmatrix}0\\1\end{pmatrix}$ respectively, then the most general unitary transformation corresponds to a $2\times 2$ matrix of the form

$$(5.1)\quad U_\vartheta \equiv \begin{pmatrix} e^{i(\delta+\sigma+\tau)}\cos(\vartheta/2) & e^{-i(\delta+\sigma-\tau)}\sin(\vartheta/2) \\ -e^{i(\delta-\sigma+\tau)}\sin(\vartheta/2) & e^{i(\delta-\sigma-\tau)}\cos(\vartheta/2) \end{pmatrix}$$

where we usually take: $\delta = \sigma = \tau = 0$.

One-bit transformations are represented as a quantum circuit:

$$(5.2)\quad |A\rangle \underline{\quad\quad\quad} U_\vartheta \underline{\quad\quad\quad} U_\vartheta|A\rangle$$

An important one-bit gate is $U_{\frac{\pi}{2}}$ which maps a spin-down particle to an equal superposition of down and up (of course, the quantum gate $U_{\frac{\pi}{2}}$ coincides with the unitary operator $U_{\sigma_2}$ in eq.(2.6)).

The vacuum state can be either $|on\rangle \equiv \begin{pmatrix}1\\0\end{pmatrix}$ or $|off\rangle \equiv \begin{pmatrix}0\\1\end{pmatrix}$

Let us suppose, for example, that the vacuum was $|on\rangle$. The action of the logic gate $U_{\frac{\pi}{2}}$ on the vacuum will be:

$$(5.3)\quad U_{\frac{\pi}{2}}|on\rangle = \frac{1}{\sqrt{2}}\begin{pmatrix}1 & 1\\-1 & 1\end{pmatrix}\begin{pmatrix}1\\0\end{pmatrix} = \frac{1}{\sqrt{2}}\left(\begin{pmatrix}1\\0\end{pmatrix} - \begin{pmatrix}0\\1\end{pmatrix}\right)$$

If instead the vacuum was $|off\rangle$:

$$(5.4)\quad U_{\frac{\pi}{2}}|off\rangle = \frac{1}{\sqrt{2}}\begin{pmatrix}1 & 1\\-1 & 1\end{pmatrix}\begin{pmatrix}0\\1\end{pmatrix} = \frac{1}{\sqrt{2}}\left(\begin{pmatrix}1\\0\end{pmatrix} + \begin{pmatrix}0\\1\end{pmatrix}\right)$$

Then, the vacuum state $|0\rangle$, after having "passed" through the quantum circuit:

$$(5.5)\quad |0\rangle \underline{\quad\quad\quad} U_{\frac{\pi}{2}} \underline{\quad\quad\quad} U_{\frac{\pi}{2}}|0\rangle$$

it comes out in the symmetric 1-qubit state:

$$(5.6) \quad U_{\frac{\pi}{2}}|0\rangle = \frac{1}{\sqrt{2}}(|on\rangle + |off\rangle) \equiv |1\rangle^S$$

or in the antisymmetric 1-qubit state:

$$(5.7) \quad U_{\frac{\pi}{2}}|0\rangle = \frac{1}{\sqrt{2}}(|on\rangle - |off\rangle) \equiv |1\rangle^A$$

Now the convenience of choosing the state $|1\rangle$ to be an equal superposition of the basis states becomes transparent.

The passage of the vacuum state through the quantum logic gate reminds us of the "quantum tunnelling" of the vacuum within the context of "Nothingness" as put forward by vilenkin [41]. However, the picture is sligthly different now: the initial geometry is not a complete "nothingness". In fact there is a pixel of area, associated with a classical bit. What is missing is time, which is unphysical. Anyway, from the quantum point of view, the initial 3-geometry may be empty (in the sense of "nothingness"), because the pixel of the boundary surface is associated with a classical bit, not with a qubit. In fact, the de Sitter horizon at the unphysical time $t = t_{-1}$ is the vacuum state:

$$(5.8) \quad |\Psi(t_{-1})\rangle = |0\rangle$$

# 6  THE "BEGINNING" OF PHYSICAL TIME

Passing through the quantum logic gate means at the same time: passing from a Boolean logic to a quantum logic, which is non-Boolean. This fact is responsible for the beginning of "physical" time. In fact, let us consider a Boolean-minded observer on the first horizon, associated with one-qubit state. He will be able to make measurements of commuting observables only in some sub-lattice of the quantum lattice. To make measurements on a different sub-lattice, he will need to move in time. The first de Sitter horizon, associated with one-qubit, is an absolute horizon as, at $t_0 = t^*$, the de Sitter horizon coincides with the horizon of a Planckian Euclidean black hole. The observer on the first horizon, cannot receive any information, because there are no preceeding horizons (for $t_{-1} = 0$ there is quantum nothingness) and succeeding horizons do not yet exist. Nevertheless, let us suppose that the observer wants to measure the area of the first horizon. He will find the area of one pixel, $A_0 = L^{*2}$. When he has completed the measurement, he pretends he is still on the first horizon; instead, he is indeed on a larger horizon.

If he will be clever enough, he could have performed the measurement in the smallest "time" available: $\Delta t = t^*$ -and now he will be on the second horizon, at time $t_1 = 2t^*$, which is associated with a 4-qubit (the area of the second horizon is of 4 pixels). So, he has got the result of one pixel when he is on a horizon whose area is of 4 pixels; 3 quantum bits of information are missing:

(6.1) $\langle \Psi(t_0) | \Psi(t_1) \rangle = \langle 1 | 4 \rangle = | 3 \rangle$

If the observer, when he was on the first horizon, had performed the measurement in a time $\Delta t = 2t^*$, at the end of the measurement, he would be on the third horizon, at time $t_2 = 3t^*$ which is associated with 9 qubits. In this case, 8 qubits of information will be missing, for the observer:

(6.2) $\langle \Psi(t_0) | \Psi(t_2) \rangle = \langle 1 | 9 \rangle = | 8 \rangle$

Let us suppose that the observer has made the measurement when he was on the second horizon, and that has taken a time $\Delta t = t^*$. At the end of the measurement, the observer will be on the third horizon, and 5 qubits of information will be missing:

(6.3) $\langle \Psi(t_1) | \Psi(t_2) \rangle = \langle 4 | 9 \rangle = | 5 \rangle$

In general:

(6.4) $\langle \Psi(t_n) | \Psi(t_m) \rangle = \langle (n+1)^2 | (m+1)^2 \rangle = | (m-n)(m+n+2) \rangle$

The number (m-n)(m+n+2) is the number of missing qubits at the end of a measurement which has taken a "time" $\Delta t = t_m - t_n = (m-n)t^*$ to be performed.

This is a case of information loss, but it is just observer-dependent. Anyway, the amount of information loss can be taken, by the observer, as a time parameter.

Let us define the information loss from slice n to slice m (with m>n) as:

(6.5) $\Delta I = (m-n)(m+n+2)$

The entropy $S_N$ of a quantum system of N qubits is:

$S_N = N \ln 2$

and the increase of entropy passing from a system of N qubits to one of M qubits is:

(6.6) $\Delta S_{MN} = (M-N)\ln 2 \geq 0$ for $M \geq N$

If we take: $N = (n+1)^2$ we have:

(6.7) $\Delta I = \Delta S_{mn} / \ln 2$

which is consistent with the Holevo-Levitin theorem [17-18], which states that the value of quantum entropy is the upper limit of the amount of quantum information one can recover from a quantum system..

In particular, for $n = -1$, we have: $S_{-1} = 0$, i.e. the vacuum has zero entropy.

Moreover, for n=0, we get: $S_0 = \ln 2$ which is the entropy of a Planckian black hole.

The amount of information loss experienced by the observer, is proportional to the increase of entropy from the $n^{th}$ de Sitter horizon to the $m^{th}$ one (with n<m).

The discrete jumps in time $\Delta t = t_m - t_n$ experienced by the observer, are related to the increase of entropy by:

(6.8) $\Delta t = \dfrac{\Delta S_{mn} t^*}{(m+n+2)\ln 2}$

and reflect the jumps from a Hilbert space $H_N$ to a Hilbert space $H_M$ (with: $N = (n+1)^2$, $M = (m+1)^2$, and m>n) performed by the discrete operator:

(6.9) $|1\rangle^{M-N}$  where:

$M - N = (m-n)(m+n+2)$.

Hence, in this model of quantum spacetime, the thermodynamic arrow of time emerges generically, and leads to a quantum cosmological theory which is time-asymmetric.

# 7 QUANTUM INFORMATION AND PRESHEAVES.

At this point, we wish to encode our model in a precise mathematical framework: the theory of categories. In particular, we shall use the theory of presheaves [29]. Presheaves are varying sets on a causal set. A causal set is a particular poset whose elements are events of a discrete spacetime.

A poset (or a partially ordered set) P is a set together with a partial order on it. Thus, a poset is formally defined as an ordered pair: $P = (S, \leq)$ where S is called the ground set of P and $\leq$ is the partial order of P. A partial order $\leq$ on a set S has:

1- Reflexivity: $a \leq a$ for all $a \in S$

2- Antisymmetry: $a \leq b$ and $b \leq a$ implies a=b

3- Transitivity: $a \leq b$ and $b \leq c$ implies $a \leq c$

A causal set C is a poset whose elements are interpreted as the events of a history.

So, for a causal set, the three properties of posets hold as well. In particular, antisymmetry is needed to avoid closed timelike loops.

A quantum causal history is a history whose events are the finite-dimensional Hilbert spaces attached to the events of the causal set.

There are discrete evolution operators acting among the Hilbert spaces associated with different causally-related events. This idea has a generic interpretation in the language of topos theory: the presheaf of Hilbert spaces over the causal set [42].

A presheaf F on a poset P is a function $F : a \rightarrow F_a$ that assignes to each $a \in P$ a set $F_a$ and to each pair $a \leq b$ a map $F_{ab} : F_a \rightarrow F_b$ such that $F_{aa} : F_a \rightarrow F_a$ is the identity map, and for $a \leq b \leq c$, the composite map is: $F_{ac} = F_{bc} \circ F_{ab}$.

A presheaf E on a causal set C will assign a Hilbert space $H_a$ to each event $a \in C$. In this case the map $E_{ab} : H_a \rightarrow H_b$ (assigned to each pair $a \leq b$) is the (discrete) evolution operator.

Remember that events that are not causally related (qubits with the same entropy) form space-like surfaces (or antichains). It should be noted that even if time is multi-fingered as in General Relativity, it happens that the slicing of spacetime can only be performed by using a lapse of time that is an integer multiple of the Planck time. A lapse of time which is not an integer multiple of the Planck time, is forbidden due to the fact that at the Planck time the corresponding qubit (which has the minimal quantum entropy) represents the quantum state of the surface horizon of a Planckian black hole. The presence of a Planckian black hole makes it impossible for the time lapse $\Delta t = t^{Planck}$ to "flow" in a continuos way.

The Alexandrov neighbourhood A (n, m) of two events n and m is the set of all x such that n<x<m. In the case of two "events" n and m=n+1, separated by a time step $\Delta t = t^{Planck}$, it happens that A (n, m) is empty. This means that, once causality is considered at the Planck scale, the notion of a Lorentzian metric becomes meaningless.

Although multi-fingered time is possible, it is not convenient. In fact, we can certainly choose to use a lapse of time $\Delta t = (m - n)t^{Planck}$ (with m>n+1), but this is a coarse-grain picture, and it leads to a larger amount of information loss than the finest-grain picture (with m=n+1) relative to the time lapse $\Delta t = t^{Planck}$. Then, it seems that in the framework of quantum information theory, there is a "preferred" slicing of spacetime, where the time lapse equals the Planck time, which leads to the minimal amount of information loss.

It appears as if one can build a discrete analogue of spacetime diffeomorphism invariance, but at the cost of a large amount of information loss.

The use of topos theory (in particular the theory of presheaves [29]) makes it possible to attach an Hilbert space to each N-qubit, and build up discrete unitary evolution operators [30], among different Hilbert spaces. In this way, we get a quantum history, which is an inflationary quantum cosmological theory.

In our model, the de Sitter horizon states $|\Psi_0\rangle, |\Psi_1\rangle, |\Psi_2\rangle, ..... |\Psi_n\rangle....$ are the events of a causal set. Perhaps we should recall that the de Sitter horizon states are N-quantum bits of information,

with $N = (n+1)^2$. The causal structure is enbodied in the quantum entropy $S = (n+1)^2 \ln 2$.

The time interval $\Delta t = t_m - t_n$ is proportional to the increase of entropy

$\Delta S = (m-n)(m+n+2)\ln 2$ from $|\Psi_n\rangle$ to $|\Psi_m\rangle$ so that $\Delta t \geq 0$ for $n \leq m$.

This gives the causal relation:

(7.1) $\quad |\Psi_n\rangle \leq |\Psi_m\rangle \quad$ for $t_n \leq t_m$.

The $|\Psi_n\rangle$ satisfy reflexivity, antisymmetry, and transitivity. In particular, transitivity implies that the causal relation is equivalent to a finite sequence of covering relations:

(7.2) $\quad |\Psi_n\rangle \to |\Psi_{n+1}\rangle \to |\Psi_{n+2}\rangle \ldots \to |\Psi_{m-1}\rangle \to |\Psi_m\rangle$

Each "event" $|\Psi_n\rangle$ is a vector of a Hilbert space $H_n = H_1^{\otimes(n+1)}$ of dimension $2^N$ where $N = (n+1)^2$ with n=0,1,2,3...

Events which are causally unrelated (in our case qubits with the same entropy) form discrete space-like surfaces (antichaines). Then, unitary evolution operators $E_{nm}$ are needed between two Hilbert spaces $H_n$ and $H_m$ associated with two causally related events $|\Psi_n\rangle$ and $|\Psi_m\rangle$.

In this way, we get a quantum history, which is an inflationary quantum cosmological theory.

The quantum history is the collection of Hilbert spaces $H_1, H_2, \ldots H_n \ldots$.

The events $|\Psi_n\rangle$ of the causal set are the de Sitter horizon states. Thus the causal set is the early inflationary universe. The quantum history is the associated cosmological theory: inflation.

The discrete evolution operators between two Hilbert spaces $H_n$ and $H_m$ are:

(7.3) $\quad E_{nm} \equiv |1\rangle^{(m-n)(m+n+2)} : H_n \to H_m$

They satisfy:

Reflexivity: $E_{nn} = 1$

Antisymmetry: $E_{nm} E_{mn} = 1$

Transitivity: $E_{mr}E_{nm} = E_{nr}$

We find an internal contradiction, in considering causality in the framework of causal sets: an observer seems to be required for consistency.

In fact, presheaves obey the Heyting algebra that implies an intuitionistic logic [31-32], which is related to the concept of time flow. One can then visualise a Boolean-minded observer who has to move in time [33] in order to grasp the underlying quantum logic of the universe. Obviously, this picture has no meaning at the Planck time, when there was no observer at all. The above picture strictly depends on the fact that in the theory of presheaves we consider Boolean sub-lattices of the quantum lattice. To drop the Boolean observer, one should discard Boolean sub-lattices and consider the entire quantum lattice as a whole. Roughly speaking, one should get an Hilbert space for the entire universe.

However, we cannot perform the tensor product of all the Hilbert spaces $H_n$ in the history, because of the presence of the unitary evolution operators $E_{nm}$ between two Hilbert spaces (although tensor product of Hilbert spaces can be performed in the consistent histories approach [43]).

In this way, we do not have a Hilbert space for the entire history (universe) as was pointed out also by Markopoulou [30].

However, we can build a (bosonic) Fock space. In fact, the evolution operators $E_{nm}$ were built up by the creation operator $a^+ \equiv |1\rangle$, which makes the number of qubits unfixed.

Then we follow the quantum statistical approach and consider the bosonic Fock space

(7.4) $\quad F(H_1) = \bigoplus_{n=1}^{\infty} H_n^{(s)}$

where $H_n^{(S)} = H_1^{(S)\otimes(n+1)}$ is a symmetric subspace of the Hilbert space $H_N$.

The logic associated with this Fock space is the direct sum of the lattice P of projectors in the various summands in the Fock space decomposition:

$P \oplus (P \otimes P) \oplus (P \otimes P \otimes P) \oplus .....$

where P denotes the lattice of projectors associated to the single-particle Hilbert space.

So, this logic is Boolean.

Nevertheless, we can get rid of the observer because the Fock space wave function leads to a atemporal description of the early universe.

In fact, the Fock space wavefunction is a linerar superposition of states $|\Psi_n\rangle = |N\rangle = |(n+1)^2\rangle$ in $H_n^{(s)}$:

(7.5) $\quad |\Psi\rangle = \sum_n \alpha_n |\Psi_n\rangle$

with the "superselection" rules:

(7.6) $\quad \langle \Psi_n | \Psi_m \rangle \equiv \langle N | M \rangle = |M - N\rangle \quad$ for N<M

$$= 0 \quad \text{for N>M}$$

$|\Psi\rangle$ is the coherent quantum state of multiple bosonic qubits and it resembles a Bose-Einstein condensate [34].

It should be noted that fermions (antisymmetric qubits) have not been taken into account as they do not contribute to the coherent quantum state.

Then, $|\Psi\rangle$ can be interpreted as the coherent quantum state of the early inflationary universe. As a Bose-Einstein condensate, $|\Psi\rangle$ will maintain coherence untill thermal noise is absent. But, at the end of inflation, when the universe starts to be radiation- dominated, $|\Psi\rangle$ will undergo decoherence by getting entangled with the environment.

## 8   THE END OF INFLATION (OR THE COLLAPSE OF $|\Psi\rangle$)

If after decoherence the environment is discarded, the surviving states behave as they had collapsed into Boolean states, and the collapse is irreversible.

This picture describes what actually occurs in a XOR gate, which is a two-qubit logic gate.

The XOR gate (or controlled-NOT gate) is the standard 2-qubits gate, and illustrates the interactions between two quantum systems.

Any quantum computation can be performed by using the XOR gate, and the set of one-qubit gates.

The XOR gate flips the "target" imput if its "control" imput is $|1\rangle$ (in our case $|on\rangle$) and does nothing if it is $|0\rangle$ (in our case $|off\rangle$):

(8.1)  $|1\rangle$ -------------------- ---------------$|1\rangle$

                    XOR             i.e      $|10\rangle \to |11\rangle$

       $|0\rangle$ ------------------ ----------------$|1\rangle$

(8.2)  $|0\rangle$ ---------------------- --------------$|0\rangle$

                    XOR            i.e      $|00\rangle$ unchanged

       $|0\rangle$ ---------------------- -------------$|0\rangle$

Hence, a XOR gate can clone Boolean imputs. But if one tries to clone a superposed state, one gets an entangled state:

(8.3)  $\frac{1}{\sqrt{2}}(|0\rangle + |1\rangle)$ ------------

                                   XOR ------------ $\frac{1}{\sqrt{2}}(|00\rangle + |11\rangle)$     (entangled state)

        $|0\rangle$ -------------------------

Then, the XOR gate cannot be used to copy superposed states (impossibility of cloning an unknown quantum state). If one of the entangled output qubits is lost in the environnement, the other qubit behaves as a classical bit, either $|0\rangle$ or $|1\rangle$.

The Fock space wave function $|\Psi\rangle$ is the coherent superposition of all the events (qubits). In this way, the atemporal picture of the early universe emerges as a coarse-graining of the theory of pre-sheaves.

Only after $|\Psi\rangle$ has undergone decoherence, one can in principle retrodict the quantum past [36], "as" it was recorded by some ancient observer.

We have given a picture of the most ancient past, but ".. the past is itself a projection of a non-Boolean structure onto our Boolean mind" [33].

The conjecture was that the beginning of inflation, (in our case we consider chaotic inflation which takes place at the Planck time $t_0 = 10^{-43} s$) was due to the passage of the vacuum through a particular one-qubit logic gate.

During inflation, the universe was vacuum dominated and the coherent quantum state in eq. (7.5) was protected against environmental noise. Thus, in the above scheme, during inflation, a collapse of the wave function could not occour.

Incidentally: One could think that during inflation, the universe behaved as a quantum computer, but it is not yet known if a subset of orthogonal unentangled states of a quantum system can perform quantum computation [44], although it has been shown that also unentangled states exibit nonlocality [45].

Let us suppose now that the coherent state $|\Psi\rangle$ collapsed at the present cosmological time:

$$T = \frac{1}{H} \approx 10^{17} s$$

where H is the Hubble constant. In Planck units, we have: $T \approx 10^{60} t^*$.

$n = 10^{60}$ would correspond to a number of qubits: $\overline{N} \approx 10^{120}$

The entropy would be at present: $\overline{S} \approx 10^{120} \ln 2$

That is about the maximal value of entropy we could have [46].

However, the actual entropy of the universe is much smaller: $S_{now} \approx 10^{101}$.

That means that the collapse of $|\Psi\rangle$ must have taken place much earlier.

The rate:

(8.4) $\overline{S}/S_{now} = 10^{19} \ln 2$

suggests that the collapse must have happened for

(8.5) $N_C \approx 10^{19}$, corresponding to:

(8.6) $n_C \approx 10^9$.

The related time is:

(8.7) $t_C = (n_C + 1)t^* \approx 10^{-34} s$.

This is just the time when inflation is supposed to end.

In fact, things changed dramatically at the end of inflation, at time $t_e \approx 10^{-34} s$, when the universe started to be radiation dominated, and quantum superpositions underwent thermal decoherence.

## 9 UNENTANGLED SYMMETRIC QUBITS AS QUANTUM HARMONIC OSCILLATORS

In Sect. 4, we have defined the creation and annihilation operators as:

$|1\rangle \equiv a^+$ ; $\langle 1| \equiv a$

with:

$aa^+ = \langle 1|1\rangle = 1$

and we have defined the projector $P_1$ as:

$P_1 \equiv a^+ a = |1\rangle\langle 1|$

In analogy with the quantum harmonic oscillator we consider the Hamiltonian:

(9.1) $H = \frac{1}{2}(aa^+ + a^+a)\hbar\omega$

In our case there will be n such Hamiltonians:

(9.2) $H_{(n)} = \frac{1}{2}(P_1 + 1)\hbar\omega_n$

one for each slice n.

The action of the operator $\frac{1}{2}(P_1 + 1)$ on the N-qubits is:

(9.3) $\frac{1}{2}(P_1 + 1)|N\rangle = |N\rangle$

so that:

(9.4) $\frac{1}{2}(P_1 + 1)|\Psi_n\rangle = |\Psi_n\rangle$

where we recall that $|\Psi_n\rangle = |N\rangle$ with $N = (n+1)^2$.

This gives:

(9.5) $H_{(n)}|\Psi_n\rangle = E_n|\Psi_n\rangle$

where $E_n = \hbar\omega_n$ ; $\omega_n = \frac{2\pi}{T_n}$ and $T_n = 2\pi(n+1)t^*$ is the period of the $n^{th}$ quantum oscillation:

Then, we identify the $n^{th}$ de Sitter horizon state with the $n^{th}$ quantum harmonic oscillator.

The discrete energy spectrum can also be found by the use of the time-energy uncertainty principle,

that, for a discrete dynamical system, is simply:

(9.6) $(E_m - E_n)(t_m - t_n) \approx \hbar$

that gives:

(9.7) $E_m - E_n \approx \frac{E^*}{m - n}$

where $E^* \approx 10^{19}$ GeV is the Planck energy.

Let us recall that $|0\rangle \equiv |\Psi_{n=-1}\rangle$ and $P_1|0\rangle = 0$.

From eq. (9.4) we have:

(9.8) $\quad H_{n=-1}|0\rangle = \frac{1}{2} E_{n=-1}|0\rangle$

If we make the assumption that at the "unphysical" time $t_{-1} = 0$ the "vacuum" energy is zero:

$E_{n=-1} = 0$, we get, from eq. (9.8):

(9.9) $\quad H_{n=-1}|\Psi_{n=-1}\rangle = 0$

and from eq. (9.7):

(9.10) $\quad E_n \approx \frac{E^*}{n+1} \quad$ (n=0,1,2,3...)

In particular, for n=0 we have $E_0 \approx E^*$ that means that the first de Sitter horizon, (the 1-qubit state) has the maximum energy, the Planck energy, as it coincides with the horizon of a Planckian black hole.

We have seen that decoherence must have occourred at time $t_c \approx 10^{-34} s = 10^9 t^*$ which is the time when inflation ends. At this time, the energy of the related N qubit state is:

(9.11) $\quad E_{n=10^9} \approx 10^{11}$ GeV $(10^{23} \,°K)$ which is the reheating temperature in this model.

The N-qubits system is completely isolated during inflation, so the superposed state can maintain coherence, but then, when inflation ends, the universe reheates by getting energy from the vacuum; at this point the qubits start to get entangled with the environment.

At the end of inflation, when $|\Psi\rangle$ decoheres, the quantum information stored in the history $H_0, H_1 ... H_{n=10^9}$ is lost to the environment.

# 10  CONCLUSIONS

We have associated a pixel (one unit of Planck area) to a quantum bit of information (instead of a classical bit as in the usual picture of holography). In this way, the main features of the holographic conjecture are unchanged but a new, quantum feature arises: n-punctured surfaces are N-qubit states with $N = (n+1)^2$.

A close relation between qubits and spin networks emerges in this picture. In fact, the surfaces which are punctured by spin network' edges in the spin-$\frac{1}{2}$ representation of SU(2), in a superposed quantum state of spin "up" and spin "down", are quantum bit states. Some issues of this picture are related to the Chern-Simons theory applied to the counting of black holes' states.

We apply this picture to the surface horizons of a de Sitter space describing a inflationary early universe. The vacuum state is a surface of one unit of Planck area associated with a classical bit of information. The vacuum undergoes a kind of quantum tunnelling by passing through a quantum logic gate, and it comes out as a one-quantum bit state. The passage of the logic gate changes the underlying logic from Boolean to non-Boolean. In this way, physical time emerges for a Boolean-minded observer. The discrete time parameter is given in terms of the discrete increase of entropy.

The one-qubit is the first de Sitter horizon at the Planck time: it coincides with the horizon of a Planckian black hole, and acts as a "creation operator" by making jumps from a Hilbert space to another one, and originating larger de Sitter horizons (inflation).

We interpret the de Sitter horizon states as events of a causal set. A presheaf of Hilbert spaces on the causal set allows us to define discrete evolution operators between two Hilbert spaces attached to two causally related events (de Sitter horizons with different area/quantum entropy).

In this way, we get a quantum history, which is the ensemble of all the Hilbert spaces.

We cannot get, however, the Hilbert space of the entire quantum history, because of the presence of the evolution operators do not allow us to make the tensor product of all the Hilbert spaces in the history. Instead, we perform the tensor sum of all the Hilbert spaces, and get a bosonic Fock space. The Fock space wavefunction resembles a Bose-Einstein condensate and leads to an atemporal description of the early universe.

We show that the de Sitter horizon states can be interpreted as quantum harmonic oscillators, and we give the discrete energy spectrum.

The coherent quantum state of qubits is completely isolated, during inflation, as the universe is vacuum-dominated, and very cool. But at the end of inflation, the universe reheates by taking energy from the vacuum, and becomes radiation-dominated. At this stage, the quantum state undergoes thermal decoherence. Once decoherence has taken place, the quantum information stored in the quantum history is lost to the environment. This fact seems to be responsible for the rather low entropy of our universe.

At this point, we wish to make some philosophical remarks.

In this model, the early inflationary universe is a superposition of quantum states (see eq. 7.5) and reminds us of the Many Worlds interpretation of quantum mechanics [47], (although this feature is lost at the end of inflation). Hence, in this context, the early universe is just a collection of "possibilities", but it is a "non being" in itself.

In the above picture, space is just the storage of the quantum information which will turn out to be necessary to build up the "real world" at the end of inflation. And time is the discrete parameter describing the increase of quantum entropy.

Thus, the "time capsules" of Barbour [48], can be visualized as the jumps of quantum entropy.

It should be noted that the whole picture is somehow related to the Spinoza's concept of "substance". A nice review on the relation between Spinoza's "substance " and quantum gravity can be found in [49].

We wish to conclude by pointing out that two different interpretations are available for this model: the atemporal one, and the causal set interpretation. While the former is related to the concept of "substance", the second one is related to the "modal perspectives of humans" (observers). Thus, it appears as if the concept of the functor Past [30], should be revisited to take into account both interpretations [50]. It should be noted, however, that holography and inflation deal only with unentangled qubits. In a more general model of space-time at the fundamental level, entangled qubits are also allowed, which is the case of a quantum computing space-time, where the very concept of causality becomes meaningless [51].

## ACKNOWLEDGMENTS

I am very grateful to Chris Isham for useful comments and remarks, in particular on the creation operator, and for illuminating discussions on the theory of presheaves.

I thank R. E. Zimmermann for useful discussions and valuable advices and comments.

I also thank Gianfranco Sartori, Roberto De Pietri, Carlo Rovelli and Lee Smolin for discussions and e-mail exchanges.

I am grateful to Francesco Lucchin for the hospitality granted at the Department of Astronomy, University of Padova, Italy.

# REFERENCES

[1] S. Mac lane and I. Moerdijk, "Sheaves in Geometry and Logic: A First Introduction to Topos Theory" (Springer-Verlag, London, 1992).

[2] R. Goldblatt, "Topoi: the categorial analysis of logic" (North-Holland, London, 1984).

[3] A. Ashtekar, "New variables for classical and quantum gravity", Phys. Rev. Lett. 57 (1986) 2244-2247; "A new Hamiltonian formulation of general relativity" Phys. Rev. D36 (1987) 1587.

[4] C. Rovelli and L. Smolin, "Knot theory and quantum gravity", Phys. Rev. Lett. 61 (1988) 1155; "Loop representation of quantum general relativity", Nucl. Phys. B133 (1990) 80-152.

[5] R. Penrose, "Theory of quantised directions" in: Quantum theory and beyond, ed. T. Bastin, (Cambridge University Press, 1971); in Advances in Twistor Theory, ed. L. P. Hughston and R.S. Word (Pitman 1979) 301.

[6] C. Rovelli and L. Smolin, "Spin networks and quantum gravity" gr-qc/9505006, Phys. Rev. D52 (1995)5743-5759.

[7] C. Rovelli and L. Smolin, "Discreteness of area and volume in quantum gravity", Nucl. Phys. B 442 (1995) 593-622.

[8] M. Reisenberg and C. Rovelli, "Sum over Surfaces form of Loop Quantum Gravity" gr-qc/9612035, Phys. Rev. D56 (1997) 3490-3508.

[9] J. C. Baez, "Spin Foam Models" gr-qc/9709052, Class. Quant. Gravity 15 (1998)1827-1858.

[10] R. Penrose, in Quantum Gravity, an Oxford Symposium, ed. C. J. Isham, R. Penrose and D. W. Sciama (Clarendon Press, Oxford 1975).

[11] L. Bombelli, J. Lee, D. Meyer and R. Sorkin, "Space-time as a causal set" Phys. Rev. Lett. 59 (1987) 521-524.

[12] F. Markopoulou and L. smolin, "Causal evolution of spin networks" Nucl. Phys. B 508 (19977) 409-430; "Quantum geometry with intrinsic local causality", gr-qc/9712067, Phys. Rev. D58 (1998)084032.

[13] G. 't Hooft, "Dimensional reduction in quantum gravity" gr-qc/9310026.

[14] L. Susskind, "The world as a hologram" hep-th/9409089, J. Math. Phys. 36(1995) 6377.

[15] J. von Neumann, "Mathematical Foundations of Quantum Mechanics" (Princeton University Press, Princeton 1955).

[16] B. Schumacher, Phys. Rev. A51 (1995) 2738.

[17] A. S. Holevo, "Coding theorems for Quantum Communication" quant-ph/9708046.

[18] L. B. Levitin, "On quantum measure of information" Proc. IV All-Union Conference on information transmission and coding theory, p. 111, Tasnkent 1969.

[19] C. E. Shannon and W. Weaver, "The mathematical theory of communication" (University of Illinois Press 1949).

[20] E. P. Specker, Dialectica 14 (1960) 175.

[21] W. K. Wooters and W. H. zureK, Nature 299 (1982) 802.

[22] D. Deutsch, Proc. R. Soc. London A 400 (1985) 97.

[23] A. Ekert, Phys. Rev. Lett. 67 (1991) 661; C. H. Bennet, g. Brassard, N. D. Mermin, Phys. Rev. Lett. 68 (1992) 557.

[24] C. H. Bennett and S. J. Wiesner, Phys. Rev. Lett. 69 (1992) 2881.

[25] C. H. Bennett, G. brassard, C. Crepeau, R. Jozsa, A. Peres and W. Wootters, "Teleporting an Unknown Quantum state via dual Classical and EPR Channels" Phys. Rev. Lett. 70 (1993) 1895.

[26] R. Landauer, "Information is physical", Physics Today, May 1991, pp.23-29.

[27] J. D. Bekenstein, Lett. Novo Cim. 11 (1974).

[28] R. Penrose, "Some remarks on Gravity and Quantum Mechanics" in Quantum Structure of Space and Time (eds. M. J. Duff and C. J. Isham, Cambridge University Press 1982).

[29] J. Butterfield and C. J. Isham " A topos perspective on the Kochen-specker theorem "(I): Quantum States as Generalized Valuations, quant-phys/9803055; (II): Conceptual aspects and classical analogues, quant-phys/ 9808067, Int. J. Theor. Phys. 38 (1999) 827.

[30] F. Markopoulou, "Quantum causal histories" hep-th/9904009; "The internal logic of causal sets: What the universe looks like from the inside" gr-qc/98011053.

[31] D. S. Bridges, S. Douglas and F. Richman, "Varieties of constructive mathematics" (Cambridge University Press, New York, 1987).

[32] "Brower's Cambridge lectures on Intuitionism" ed D. Van DAlen (cambridge University Press,1981).

[33] A. A. Grib, in Quantum Cosmology and the Laws of Nature, eds. R. J. Russel, N. Murphy and C. J. Isham; A. A. Grib and R. R. Zapatrin, Int. J. Theor. Phys. 29 (1990) 113.

[34] M. H. Anderson, J. R. Ensher, M. R. Matthews, C. E. Wietman and E. A. Cornell, "First Observation of a Bose-Einstein Condensate", Science 296 (1995) 198.

[35] J. J. Halliwell, "Somewhere in the Universe: where is the Information Stored When Histories Decohere?" quant-ph/9902008, Phys. Rev. D 60 (1999) 904-907.

[36] J. B. Hartle, "Quantum Pasts and the Utility of History" gr-qc/9712001.

[37] A. Ashtekar, J. Baez, A. Corichi, K. Krasnov, "Quantum Geometry and Black Hole Entropy" gr-qc/9710007, Phys. Rev. Lett. 80 (1998) 904; A. Ashtekar and K. Krasnov, "Quantum Geometry and Black Holes" gr-qc/9804039.

[38] G. Immirzi, "Quantum gravity and Regge calculus" gr-qc/9701052, Nucl. Phys.Proc. Suppl. 57 (1997) 65; C. Rovelli and T. Thiemann, "The Immirzi parameter in quantum general relativity" gr-qc/9705059, Phys. Rev. D57 (1998) 1009.

[39] J. Wheeler, in Sakharov Memorial Lectures on Physics, vol.2, edited by L. Keldysh and V. Feinberg (Nova Science, New York, 1992).

[40] P. A. Zizzi, "Quantum Foam and de Sitter-like Universe" hep-th/9808180, IJTP Vol. 38, N. 9, (1999) 2333-2348.

[41] A. Vilenkin, "Creation of Universes from Nothing", Phys. Lett.117B (1982) 25.

[42] C. J. Isham, private communication.

[43] C. J. isham, "Topos Theory and Consistent Histories: The Internal Logic of the Set of all Consistent Sets" gr-qc/9607069, IJTP 36 (1997) 785-814.

[44] C. H. Bennett, D. P. Di Vincenzo, C. A. Fuchs, T. Mor, E. Rains, P. W. Shor, J. A. Smolin and W. K. Wootters, "Quantum Nonlocality without Entanglement" quant-ph/9804053.

[45] J. A. Smolin, private communication.

[46] R. Penrose, The emperor's New Mind (Oxford University Press 1989).

[47] H. Everett III, "Relative State" Formulation of Quantum Mechanics, Rev. of Modern Physics Vol. 29, (1957) 454.

[48] J. Barbour, The End of Time (The Next Revolution in our Understanding of the Universe), Weidenfeld & Nicolson, London, 1999.

[49] R. E. Zimmermann, "Loops, and Knots as Topoi of substance. Spinoza Revisited", submitted to: Foundations of Physics.

[50] R. E. Zimmermann and P. A. Zizzi, "The Functor Past Revisited. A quantum Computational Model of Emergent Consciousness", work in progress.

[51] P. A. Zizzi, "Quantum Computing Space-Time", to appear.41